# Long time-series NDVI reconstruction in cloud-prone regions via spatio-temporal tensor completion

Dong Chu [a], Huanfeng Shen [a,b], Xiaobin Guan [a,c,*], Jing M. Chen [c,d], Xinghua Li [e], Jie Li [f], Liangpei Zhang [b,g]

[a] *School of Resource and Environmental Sciences, Wuhan University, Wuhan 430079, P.R. China.*

[b] *Collaborative Innovation Center of Geospatial Technology, Wuhan 430079, PR China.*

[c] *Department of Geography and Planning, University of Toronto, Toronto ON M5S3G3, Canada.*

[d] *School of Geographical Sciences, Fujian Normal University, Fuzhou 350117, China.*

[e] *School of Remote Sensing and Information Engineering, Wuhan University, Wuhan 430079, China*

[f] *School of Geodesy and Geomatics, Wuhan University, Wuhan 430079, China*

[g] *The State Key Laboratory of Information Engineering in Surveying, Mapping and Remote Sensing, Wuhan University, Wuhan 430079, China*

\* Corresponding author: Xiaobin Guan (guanxb@whu.edu.cn)

## Abstract

The applications of Normalized Difference Vegetation Index (NDVI) time-series data are inevitably hampered by cloud-induced gaps and noise. Although numerous reconstruction methods have been developed, they have not effectively addressed the issues associated with large gaps in the time series over cloudy and rainy regions, due to the insufficient utilization of the spatial and temporal correlations. In this paper, an adaptive Spatio-Temporal Tensor Completion method (termed ST-Tensor) method is proposed to reconstruct long-term NDVI time series in cloud-prone regions, by making full use of the multi-dimensional spatio-temporal information simultaneously. For this purpose, a highly-correlated tensor is built by considering the correlations among the spatial neighbors, inter-annual variations, and periodic characteristics, in order to reconstruct the missing information via an adaptive-weighted low-rank tensor completion model. An iterative $\ell_1$ trend filtering method is then implemented to eliminate the residual temporal noise. This new method was tested using MODIS 16-day composite NDVI products from 2001 to 2018 obtained in the region of Mainland Southeast Asia, where the rainy climate commonly induces large gaps and noise in the data. The qualitative and quantitative results indicate that the ST-Tensor method is more effective than the five previous methods in addressing the different missing data problems, especially the temporally continuous gaps and spatio-temporally continuous gaps. It is also shown that the ST-Tensor method performs better than the other methods in tracking NDVI seasonal trajectories, and is therefore a superior option for generating high-quality long-term NDVI time series for cloud-prone regions.

## 1. Introduction

The Normalized Difference Vegetation Index (NDVI), as obtained by optical satellites is one of the most popular indicators for the status of vegetation (Myneni and Williams, 1994). NDVI data have therefore been widely utilized in terrestrial carbon cycle modeling(Guan et al., 2019, 2018; Tucker and Sellers, 1986), vegetation phenology detection (Cao et al., 2015; Piao et al., 2019; Zhang et al., 2003), and land-cover change monitoring (Lunetta et al., 2006; Tang et al., 2019). However, satellite-derived NDVI data are inevitably contaminated by clouds, aerosols, and other adverse atmospheric conditions, which greatly hinder its further application (Huete et al., 2002). Therefore, to represent the actual conditions of vegetation, it is of great importance to reconstruct the contaminated information and obtain a high-quality NDVI time series (Shao et al., 2016).

In the past decades, a number of temporal reconstruction methods have been proposed for generating high-quality NDVI time series, and have become widely used due to their advantages of simplicity and high efficiency. Generally speaking, these methods can be divided into three main types (Shen et al., 2015). The first type of method is the function-based curve fitting methods, which fit the entire time-series NDVI curve by a mathematical function, such as an asymmetric Gaussian (AG) function (Jönsson and Eklundh, 2002), double logistic (DL) function (Beck et al., 2006), or cubic spline polynomial (Chen et al., 2006). This category of methods can obtain a relatively smooth result with robust seasonal dynamics, but such methods usually fail to capture the short-term changes in vegetation growth (Zhu et al., 2012). The second type of method is the sliding window filtering methods, which process the NDVI time series in a local moving window. Examples of sliding window filter methods are the best index slope extraction (BISE) algorithm (Viovy et al., 1992), the iterative interpolation for data reconstruction (IDR) method (Julien and Sobrino, 2010), the Savitzky-Golay (SG) filter (Chen et al., 2004), and the adapted local regression filter (Moreno et al., 2014). However, for these methods, the selection of filtering parameters greatly affects the filtering results. The third type of method refers to the frequency domain approaches, such as the harmonic analysis of time series (HANTS) method (Yang et al., 2015; Zhou et al., 2015) and the wavelet method (Lu et al., 2007). However, it is also difficult to determine the parameters for these methods. There are also some methods that cannot be categorized into these three main types, such as the data assimilation method (Gu et al., 2009) and the variation-based filtering methods (Atzberger and Eilers, 2011; Kong et al., 2019). For example, the Whittaker filtering method is a typical variation-based filtering method that pursues a balance between fidelity and smoothness, based on regularization theory (Kim et al., 2009).

However, these temporal methods invariably treat every pixel individually, and only the temporal information of the pixel itself is used for the reconstruction. This limitation leads to a poor performance in the situation of large time gaps. In view of this, some attempts have been made to utilize spatial domain information to assist with the temporal filtering, and spatio-temporal methods have been proposed. For example, Gerber et al. (2018) and De Oliveira et al. (2014) attempted to make use of the essential information obtained from a spatio-temporal window; Verger et al. (2013) and Xu et al. (2015) made attempts to determine contaminated pixels by the use of the high-quality pixels from the same land-cover type. However, these two methods generally require additional data, e.g., a land-cover map or an ecological zone map. Recently, Cao et al. (2018) proposed the Spatial-Temporal Savitzky-Golay (STSG) method, which assumes that clouds are spatially discontinuous, and employs neighboring pixels to assist with the noise reduction of the target pixel. The results obtained by these methods have proved that spatial information can help to deal with the continuous gaps in the NDVI time series.

Nevertheless, all of the previous methods are still unable to adequately mine and utilize the temporal and spatial reference information when filling gaps in the NDVI time series. For the temporal domain, almost all of the temporal methods tend to prefill cloud-contaminated pixels by simple temporal linear interpolation, which only uses the temporal information of temporally neighboring observations, and ignores the periodic characteristics of vegetation in the accumulated decades of the NDVI time series. For the spatial domain, the current spatio-temporal methods place emphasis on reconstructing missing pixels by the detected similar pixels in neighboring pixels, and they fail to achieve collaborative use of the spatio-temporal information. In fact, long-term NDVI time series can provide abundant reference information for gap-filling, from the spatial neighbors, temporal neighbors, and periodic characteristics. However, none of the current methods can make full use of the three-dimensional information simultaneously. These deficiencies limit the accuracy of the current methods in cloud-prone areas, especially in tropical and monsoon regions with a rainy climate, where temporally continuous gaps or even spatio-temporally continuous gaps frequently exist in the data (Liu et al., 2017; Zhou et al., 2016).

In this paper, a novel spatio-temporal adaptive tensor completion (ST-Tensor) method is proposed to reconstruct the severe data gaps in long-term NDVI products over cloud-prone areas. The core idea of the proposed method is to treat the time-series NDVI data as a tensor, which is a high-order generalization of vectors and matrices (Kolda and Bader, 2009), so as to deliver the intrinsic multi-dimensional spatio-temporal correlations fully and equally. In detail, the ST-Tensor method first constructs a low-rank tensor by synthetically utilizing the information from the multi-dimensional spatio-temporal domain, including the spatial neighbors, inter-annual variations, and periodic characteristics. An adaptive low-rank tensor completion method is then used to generate gapless time-series NDVI data.

In the rest of this paper, a detailed description of the proposed method is first provided. We then describe the qualitative and quantitative comparison experiments conducted in this study. Finally, a discussion on the proposed method are presented.

## 2. Study area and data

In this research, Mainland Southeast Asia was selected as the study area, which is characterized by tropical humid climate in the coastal areas and a monsoon climate in the interior areas, and is known as a typical cloud-prone region (Dong et al., 2012; Leinenkugel et al., 2013b, 2013a; Xiao et al., 2006). Mainland Southeast Asia covers approximately 2.05 million square kilometers, including Vietnam, Burma/Myanmar, Laos, Thailand, and Cambodia. The region is mainly covered by tropical evergreen forest and farmland and has been identified as a biodiversity hotspot (Sloan et al., 2014).

MOD13A2 version 6 NDVI data (Didan et al., 2015) were collected in this study for testing the proposed reconstruction method and the other four comparison methods. The spatial resolution of the MOD13A2 product is 1 km, and the temporal resolution is 16 days. We collected the MOD13A2 version 6 NDVI data for the study area from 2001 to 2018 from the website of the National Aeronautics and Space Administration

([https://search.earthdata.nasa.gov/search](https://search.earthdata.nasa.gov/search)). Although the maximum value composite algorithm has been applied to the data, cloud-contaminated pixels still exist in the product. The product includes a summary quality layer that records the pixel reliability index (RI) of every NDVI value, according to which we can establish whether a pixel is contaminated or not. In this study, an RI value of −1 (no data) or 3 (cloudy) was considered as a flag for invalid pixels that needed to be filled, and an RI value of 0 (good data) or 1 (marginal data) was considered as a flag for valid pixels that could be used for filling. Fig. 1 shows the invalid data percentage of every pixel in the study area over the whole time span from 2001 to 2018. The invalid percentage is equal to the number of invalid observations divided by the total number of observations in the time series. As can be seen, most of the study area has an invalid percentage of more than 20%, and the mean invalid percentage reaches 24%. The high percentage of invalid observations poses a great challenge to the reconstruction of the NDVI time series.

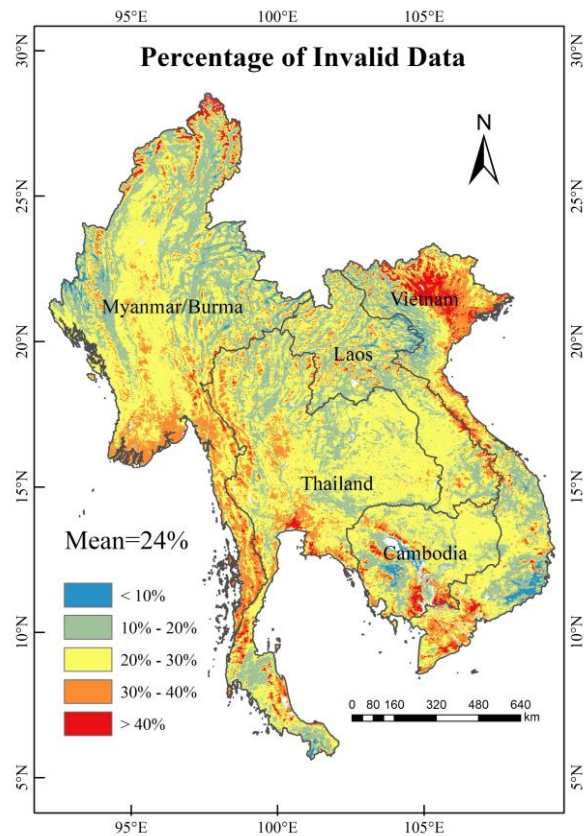

**Fig. 1.** Invalid data percentage of every pixel location in the study area over the whole time span from 2001

to 2018.

## 3. Method

### 3.1 Spatio-temporal adaptive tensor completion model

A vector is one-dimensional, a matrix is two-dimensional, and a tensor is more than two-dimensional and is often regarded as a high-order generalization of vectors and matrices (Kolda and Bader, 2009). In this paper, we represent tensors by bold capital letters, e.g., $\mathbf{X}$, and matrices are denoted by non-bold capital letters, e.g., $X$. An $N$-order tensor $\mathbf{X} \in \mathbb{R}^{I_1 \times I_2 \times ... \times I_N}$ can be unfolded into $N$ matrices from $N$ different modes. The mode-n unfolding of $\mathbf{X}$ is denoted as $X_n \in \mathbb{R}^{I_n \times \prod_{\substack{j=1 \\ j \neq n}}^{N} I_j}$. The tensor's rank can characterize the correlations in the different domains. Based on these N unfoldings, the rank of tensor $\mathbf{X}$ can be defined as a rank array of the unfolding matrices(Gandy et al., 2011; Kolda and Bader, 2009; Liu et al., 2013). Tensor $\mathbf{X}$ is considered to be of low-rank in the case of every $X_{(n)}$ being of low rank.

For a small patch of the original long time-series NDVI cube, we denote it as a third-order tensor $\mathbf{Y} \in \mathbb{R}^{m \times m \times T}$, where $m$ represents the side length of the block, i.e., the spatial dimension, and $T$ signifies the number of time-series images, i.e., the temporal dimension. Correspondingly, an index set $\Omega$ can also be obtained based on the RI values. If the position $(i, j)$ in the $t$ th image is missing or contaminated (i.e., the RI value is −1 or 3), $\Omega_{i,j,t} = 0$. If the position is a clean-sky pixel or marginal pixel (i.e., the RI value is 0 or 1), $\Omega_{i,j,t} = 1$. An ideal complete tensor $\mathbf{X} \in \mathbb{R}^{m \times m \times T}$ needs to be recovered under the assumption that $\mathbf{X}$ is low rank. Such a problem can be expressed as the following optimization problem:

$$\min_{\mathbf{X}} \quad \text{rank}(\mathbf{X}) \quad \text{s.t.} \quad \mathbf{X}_\Omega = \mathbf{Y}_\Omega \tag{1}$$

The key to solving this problem is how to approximate the rank of the tensor. Based on the definition of the n-rank of a N-dimensional tensor, the following model (Eq. (2)) can be obtained, and the original problem is transformed into a problem of approximation of the matrix rank.

$$\min_{\mathbf{X}} \sum_{n=1}^{3} w_n \operatorname{rank}(X_{(n)}) \quad \text{s.t.} \quad \mathbf{X}_\Omega = \mathbf{Y}_\Omega \tag{2}$$

where $w_n$ is the weight corresponding to $X_{(n)}$, which is always non-negative and satisfies $\sum_{n=1}^{3} w_n = 1$. In order to achieve satisfactory completion results, the weight $w_n$ is adaptively determined according to the low-rank degree of the different dimensions of the tensor, as further described in Section 3.3.

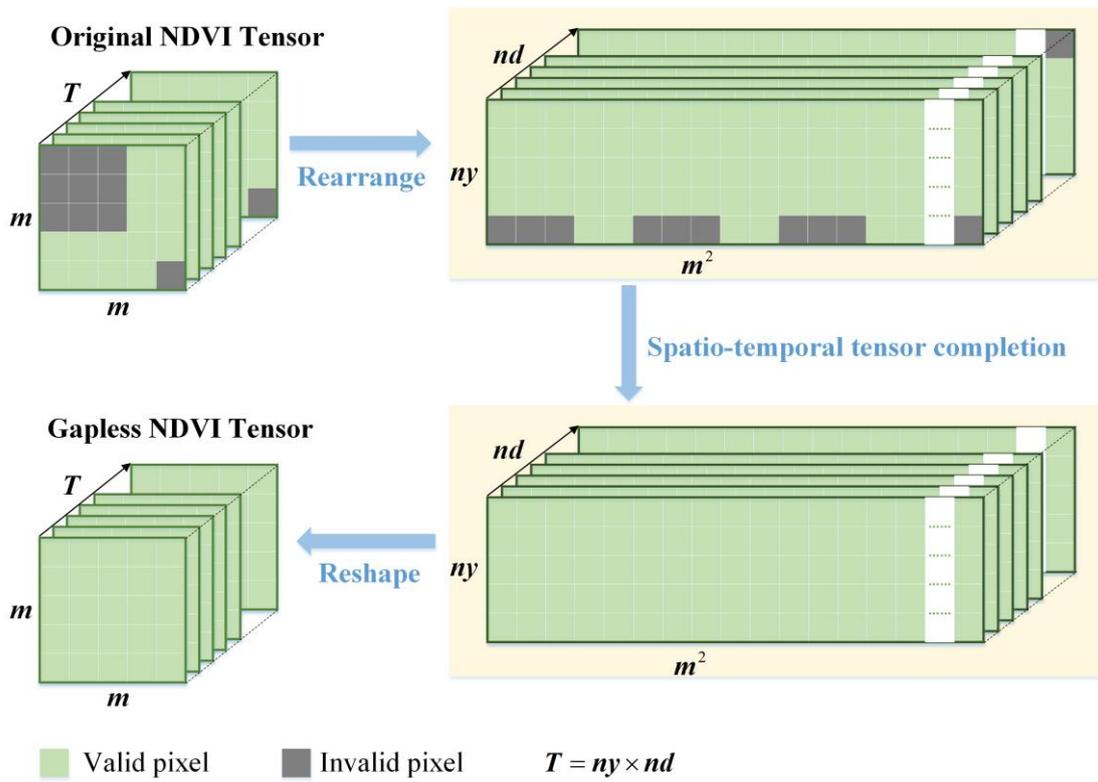

**Fig.2.** Graphical illustration of the long time-series NDVI gap-filling by spatio-temporal tensor completion. $m$ denotes the side length of the spatial patch; $T$ represents the total number of observations in the whole time series; $ny$ is the number of years; and $nd$ is the number of observations in a year, which is 23 in this study. $T$ is equal to $ny$ multiplied by $nd$.

## 3.2 Tensor rearrangement

Although spatio-temporal tensor completion can be conducted on the original third-order tensor form,

i.e., the optimization problem in Eq. (3), the spatio-temporal correlations cannot be fully leveraged. In essence, the low-rank degree of the tensor, i.e., the similarity degree of each dimension of the tensor, determines the accuracy of the low-rank tensor reconstruction. In this study, the low-rank degree of the tensor can be further strengthened by tensor rearrangement, due to the three obvious characteristics of time-series NDVI images. The first characteristic is the spatial neighborhood similarity, which means that a pixel will be similar to its spatially neighboring pixels. This characteristic is built on the first law of geography. The second characteristic is the temporal neighborhood correlation, which means that the $t$th image will depend on its temporally neighboring images. This characteristic is based on the assumption that vegetation growth is continuous and smooth. The third characteristic is the periodic temporal similarity, which means that the NDVI values on the same day of a year but in different years will be similar. This characteristic is built on the fact that vegetation growth changes periodically. It is clear that the original tensor form can express the spatial neighborhood similarity and temporal neighborhood correlation. The temporal periodicity, however, is not directly expressed in this form. For this reason, it is necessary to reshape the original tensor into a more reasonable form that can take these three characteristics into consideration.

As shown in Fig. 2, the core idea of reshaping is to turn the one-dimensional multi-year time-series of each pixel into a two-dimensional matrix, of which each column is the time series of one year, and to turn the two-dimensional spatial patch into a vector. Thus, the periodicity can be expressed explicitly. In detail, for the original tensor $\mathbf{Y} \in \mathbb{R}^{m \times m \times T}$, where $m$ denotes the side length of the spatial patch and $T$ represents the total number of observations in the whole time series, which is essentially equal to the number of years, i.e., $ny$, multiplied by the number of observations in a year, i.e., $nd$, which was 23 in this study. After rearrangement, this block become a new three-dimensional block that is denoted as $\mathbf{Y}' \in \mathbb{R}^{m^2 \times nd \times ny}$. The corresponding index set $\Omega'$ is reshaped into $\Omega$ in the same way. After tensor rearrangement, we can derive the following model according to Eq. (3):

$$\min_{\mathbf{X}'} \sum_{n=1}^{3} w_n \operatorname{rank}(X'_{(n)}) \quad \text{s.t.} \quad \mathbf{X}'_{\Omega'} = \mathbf{Y}'_{\Omega'} \tag{3}$$

Eq. (3) can be solved by the algorithm proposed by Ji et al. (2017).

**3.3 Iterative weight updating**

Different unfolding matrices should be given different weights according to their low-rank degree. Specifically, the $n$ th unfolding matrix $X'_{(n)}$ should be given a larger (or smaller) weight $w_n$ if it has a stronger (or weaker) low-rank property. However, the question remains of how to measure the low-rank property for different matrices uniformly? It is well known that the faster the singular values of a matrix decrease, the stronger the low-rank property of the matrix. From this idea, the weight can be calculated based on the distribution of the singular values of the matrix. A detailed description is provided in the following.

Firstly, the singular values of the matrix $\hat{X}_{(n)}$ can be obtained through singular value decomposition (SVD). They are then sorted in descending order and denoted as a vector:

$$\sigma^{(n)} = [\sigma_1^{(n)}, \sigma_2^{(n)}, \ldots, \sigma_{s_n}^{(n)}] \tag{4}$$

where $s_n$ is the number of singular values of matrix $\hat{X}_{(n)}$.

Secondly, a parameter $\alpha$ that signifies the information proportion of the first $k$ singular values in all $s_n$ singular values is defined as:

$$\alpha = \frac{\sum_{i=1}^{k} \sigma_i^{(n)}}{\sum_{i=1}^{s_n} \sigma_i^{(n)}} \tag{5}$$

For a constant $k$, the stronger the low-rank property of the matrix, the bigger the value of $\alpha$ is. Conversely, if we give a constant $\lambda$, then the stronger the low-rank property of the matrix, the smaller the value of $k$. In this study, we set a threshold of $T=0.85$. For the different unfolding matrices $\hat{X}_{(n)}$, we find the corresponding $k_n$ that is defined as the smallest $k$ when $\lambda \geq T$. $k_1$, $k_2$, and $k_3$ for $\hat{X}_{(1)}, \hat{X}_{(2)}$, and $\hat{X}_{(3)}$ can then be obtained, respectively.

Thirdly, it is necessary to normalize $k_n$ to the same scale. Therefore, the normalized value is defined as:

$$\tilde{k}_n = \frac{k_n}{s_n} \quad (6)$$

The different $\tilde{k}_n$ now have a universal scale, and can be used to calculate the weight. As mentioned earlier, the smaller the value of $\tilde{k}_n$, the stronger the low rank property of the corresponding matrix $\hat{X}_{(n)}$, and a bigger weight $w_n$ should be given to make the recovery result of this matrix contribute more to the final result. Because of the inverse relationship between $\tilde{k}_n$ and $w_n$, and at the same time, $w_n$ has to satisfy $\sum_{n=1}^{3} w_n = 1$, a simple expression is defined for $w_n$:

$$w_n = \frac{\frac{1}{\tilde{k}_n}}{\frac{1}{\tilde{k}_1} + \frac{1}{\tilde{k}_2} + \frac{1}{\tilde{k}_3}} \quad (7)$$

. Initially we initialize the weights of the three unfolding matrices to be equal, i.e., $w_1 = w_2 = w_3 = \frac{1}{3}$. Then, in each iteration, $w_n$ should be updated adaptively according to the description.

### 3.4 Iterative $\ell_1$ trend filtering

Although a gap-free NDVI time series can be obtained after the processing of the spatio-temporal adaptive tensor completion, there may still be some noise in the data due to the uncertainties in the RI data. Due to the flexibility of the variation-based methods to denoise one-dimensional time-series data by regularizing the residual term and the smoothness term, we propose an iterative $\ell_1$ trend filtering method to further filter the time-series NDVI.

A noisy time series is denoted as $y$ and the fitted series as $z$. The $\ell_1$ trend filtering method attempts to balance two conflicting goals: (a) fidelity to the original series and (b) smoothness of the filtered series (Eilers, 2003). It obtains the following objective function (Eq. (8)). $D$ is a second-order difference matrix.

$\lambda$ is the regularization parameter used to balance the fidelity term and the smoothness term.

$$Q = \frac{1}{2}\|y - z\|_2^2 + \lambda\|Dz\|_1 \tag{9}$$

In practical NDVI time-series filtering applications, due to the nature of the $\ell_1$ norm constraint, $\ell_1$ trend filtering has a better ability to maintain the detailed characteristics of the turning points compared to the Whittaker filter. We propose an iterative procedure based on the general assumption that noise is negatively biased in the NDVI time series. Even after the process of spatio-temporal adaptive tensor completion, the remaining noise can also be assumed to be negatively biased. We treat the marginal data (i.e., RI=1) and reconstructed data as noisy, while the good data (i.e., RI=0) is considered to be almost noise-free. The iterative procedure follows the following process. In the first and second runs, the basic $\ell_1$ trend filtering is made to smooth the NDVI time series, so that the good data remain unchanged, and only the noisy values that lie below the smoothed series are replaced. For the final run, the basic $\ell_1$ trend filtering is implemented without the replacement, and as the output result. Through such an iterative procedure, a good balance can be achieved between noise reduction and good data preservation. After the filtering step is completed, the final results, which are not only gap-free but also noise-free, can be acquired.

## 4. Experiments and results

In the experiments, the proposed method was compared with four temporal filtering methods and one spatio-temporal method, i.e., the HANTS method (Zhou et al., 2015), the IDR method (Julien and Sobrino, 2010), the Whittaker filter method (Atzberger and Eilers, 2011), the SG filter method (Chen et al., 2004), and the STSG method (Cao et al., 2018). It is worth noting that time domain linear interpolation was implemented to fill contaminated pixels before applying the time-series filtering methods. We reconstructed long time-series NDVI images in the study area for 2001–2018 using the proposed method and the comparison methods.

### 4.1 Assessments from the curves of representative points

To demonstrate the performance of the ST-Tensor method in reconstructing NDVI time-series, 10 typical vegetation pixels were selected based on the land-cover types for the comparison between the original and the reconstructed time series of the different methods. In order to facilitate visual evaluation, three of the five comparison methods, i.e., the Whittaker filter method, the SG filter method, and the STSG method, are selected for display. Fig. 3 shows the long-term results of these three methods and the proposed ST-Tensor method for three vegetation pixels. The entire 18-year time series is presented to assess the overall performance in a long time series. In general, the ST-Tensor method exhibits a better and more stable performance across all 18 years, in all cases, not only in gap-filling but also in noise-reduction. Magnified details are shown in Fig. 4, which shows a specific comparative analysis in zoomed temporal windows.

The main advantage of the ST-Tensor method over the other methods is its robust ability to restore gaps, for the isolated gaps, continuous gaps, and gaps at the peaks and valleys, due to the integrated utilization of the spatio-temporal information in all years. The temporal filtering methods, i.e., the Whittaker and SG methods, are able to address the isolated gaps (Fig. 4(j)) and narrow gaps (Fig. 4(f)) in the vegetation growth periods, but they fail to deal with the gaps in the peak growth season as shown in Fig. 4(a), Fig. 4(g), and Fig. 4(h), as well as the continuous gaps in Fig. 4(a), Fig. 4(c), Fig. 4(d), and Fig. 4(e). The STSG method is superior to the temporal methods in almost all cases, because of the use of spatial information, which is basically in line with the results presented in Cao et al. (2018). However, the STSG method shows some instability at some points in the time series. An obvious example can be observed in Fig. 4(d) and Fig. 4(e), where the two pixels exactly represent double-season rice. Due to the continuous cloud contamination in the summer when the first-season-rice harvest occurred, there is no essential information in both the spatial and temporal neighbors. In this condition, STSG fails to reconstruct the real valleys in some years, i.e., 2014 in Fig. 4(d) and 2018 in Fig. 4(d), and only the ST-Tensor method successfully restores the valley characteristics in all years because it can use the periodical temporal information from the other years.

Relative to the other methods, the ST-Tensor method has another advantage in dealing with negatively-biased marginal values and in eliminating slightly abnormal fluctuations. Taking Fig. 4(f) as an example, there is some marginal noise and frequent temporal fluctuations in the original time series. All of the methods can dampen the abnormal fluctuations to some extent and obtain a smoother curve. However, if the marginal data have a relatively large negative deviation, the reconstruction results of the Whittaker method are slightly poorer. Furthermore, the Whittaker method overestimates the normal valleys in nearly all years. In this case, the STSG and ST-Tensor methods can address the problem of marginal data with low values much better, but the result of the STSG method still shows some minor fluctuations, especially in the peak growing season, which is possibly caused by the parameters of the SG filter.

The ST-Tensor method also shows a stable simulation ability in both the peak and valley periods. The Whittaker method does not perform well in these periods due to the over-smoothing (as can be seen in Fig. 4(d) and Fig. 4(e)). The SG and STSG methods tend to reach the upper envelope of all the values, whereas the ST-Tensor method tends to reach the upper envelope of the good values, instead of all the values (Fig. 3(a)). This small difference can sometimes prevent over-estimation in the peak periods.

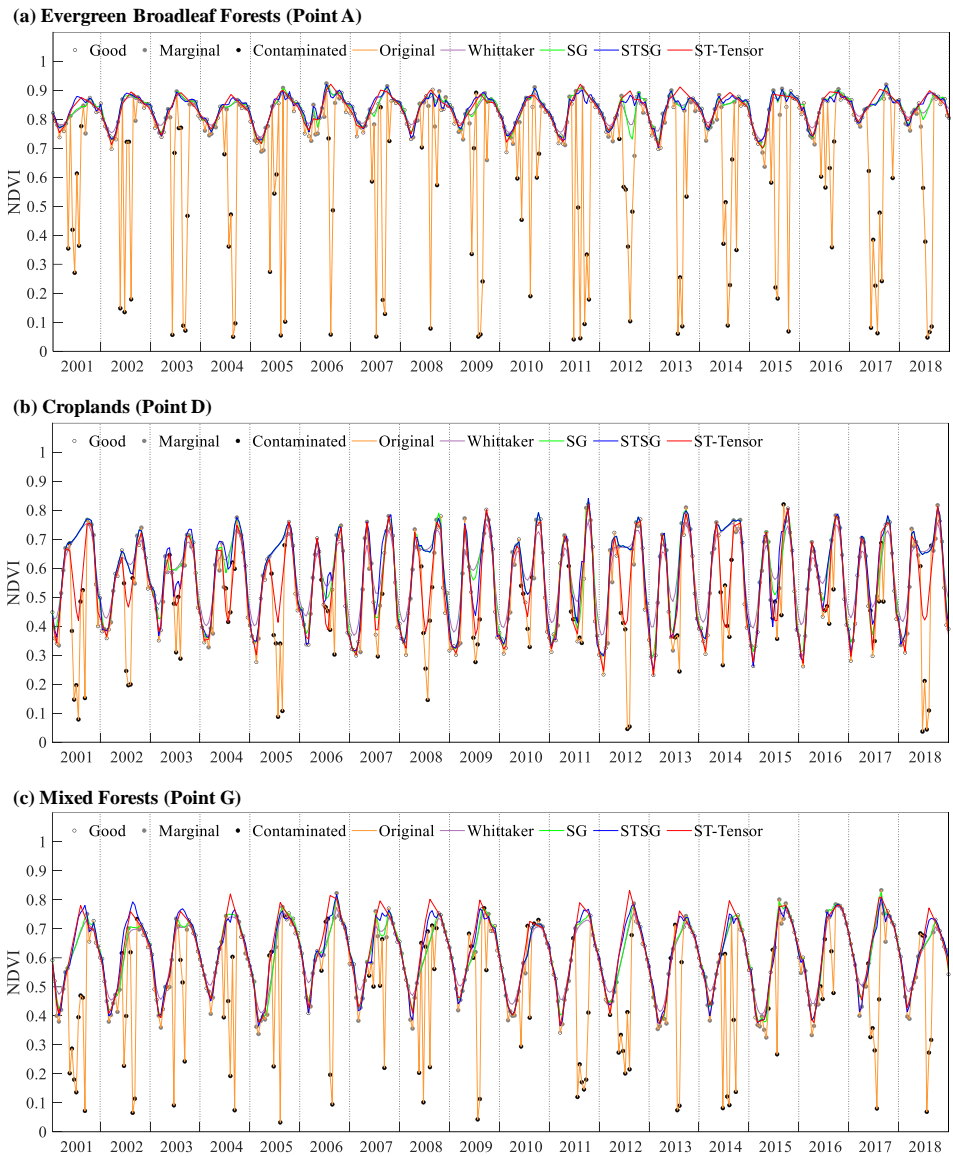

**Fig. 3.** The full 18-year comparisons for the results of the proposed ST-Tensor method and the other three methods for three typical vegetation pixels.

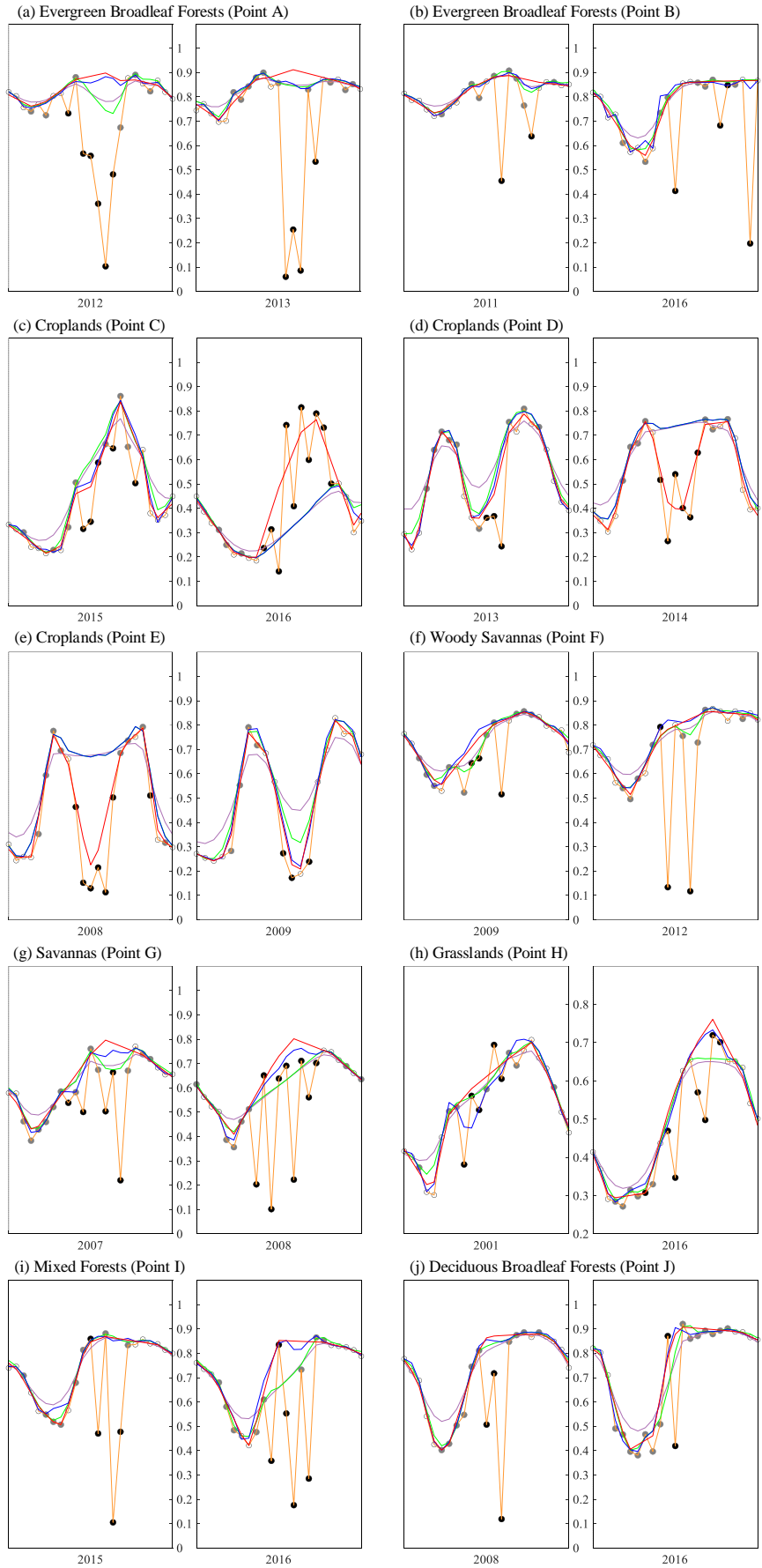

**Fig. 4.** Comparisons of the results of the proposed ST-Tensor method and the other three methods for 10 typical vegetation pixels.

**4.2 Assessments from the spatial distribution perspective**

In this section, the reconstruction results of the ST-Tensor method and the other three methods in Mainland Southeast Asia are compared from the perspective of the spatial distribution. Fig. 5 presents the spatial distribution of the proposed ST-Tensor method for the whole of Mainland Southeast Asia on DOY 241 in 2004, which is in the summer season with high vegetation coverage and serve cloud contamination. From a spatial perspective, the ST-Tensor method can increase the contaminated low NDVI values up to normal high values and obtain a reconstruction result that is spatially continuous. Fig. 6 shows the detailed performance of the different methods in two local areas. Fig. 6(a) shows a region with high NDVI values, which is degraded by severe cloud cover. Although the low values can be corrected by the Whittaker filter and SG filter, there is still obvious negatively-biased noise and spatial discontinuities in the results. The STSG method performs better, but there is still some noise in the results. The proposed ST-Tensor method shows the best performance in terms of the noise reduction and spatial continuity. Fig. 6(b) shows a region that is mainly cropland, where the date is close to the date of the peak of the crop growth. The proposed ST-Tensor method and STSG method restore the low NDVI values successfully, whereas both the Whittaker method and SG method underestimate the NDVI values. Overall, the results demonstrate the good performance of the ST-Tensor method in keeping key points from a spatial perspective.

As shown in Fig. 1, northern Vietnam shows a much higher invalid percentage, which poses a great challenge to the reconstruction of the NDVI time-series. Comparisons between the raw and reconstructed NDVI in this region for the same day in two different years are presented in Fig. 7(a) and Fig. 7(b). Fig. 7(a) corresponds to DOY 17 in 2010 and Fig. 6(b) corresponds to DOY 17 in 2012. There are a few missing and noisy pixels for DOY 17 in 2010, for which the NDVI values are relatively low. It can be observed that most of the low NDVI values are successfully increased by all the methods, and the reconstruction results of the different methods are almost the same. However, there is still a region with low NDVI values. In fact, the main land cover type in this region is cropland, which had been harvested in January. Therefore, these low NDVI values are in fact accurate.

Fig. 7(b) shows the results for DOY 17 of 2012, which has large spatially continuous gaps. The ST-Tensor method, Whittaker method, and SG method increase the contaminated low NDVI values up to high values while the STSG method fails to reconstruct some pixels and leads to an abnormal result. This suggests that the STSG method still has some problems in the case of spatially continuous gaps. When compared with Fig. 7(a), it can be found that the reconstruction results of all the temporal-based methods in the cropland are incorrectly high. Only the proposed ST-Tensor method can restore the NDVI values in this area. Actually, for this region, spatially continuous missing pixels exist not only on DOY 17, but also on DOY 1, DOY 33, DOY 49, and DOY 65 of 2012 (Fig. 7(c)). The temporally continuous missing data lead to the poor performance of these temporal-based methods. The STSG method also fails to deal with the situation of spatio-temporally continuous missing data. Fig. 7(c) shows that the proposed ST-Tensor method can successfully reconstruct spatio-temporally continuous missing pixels.

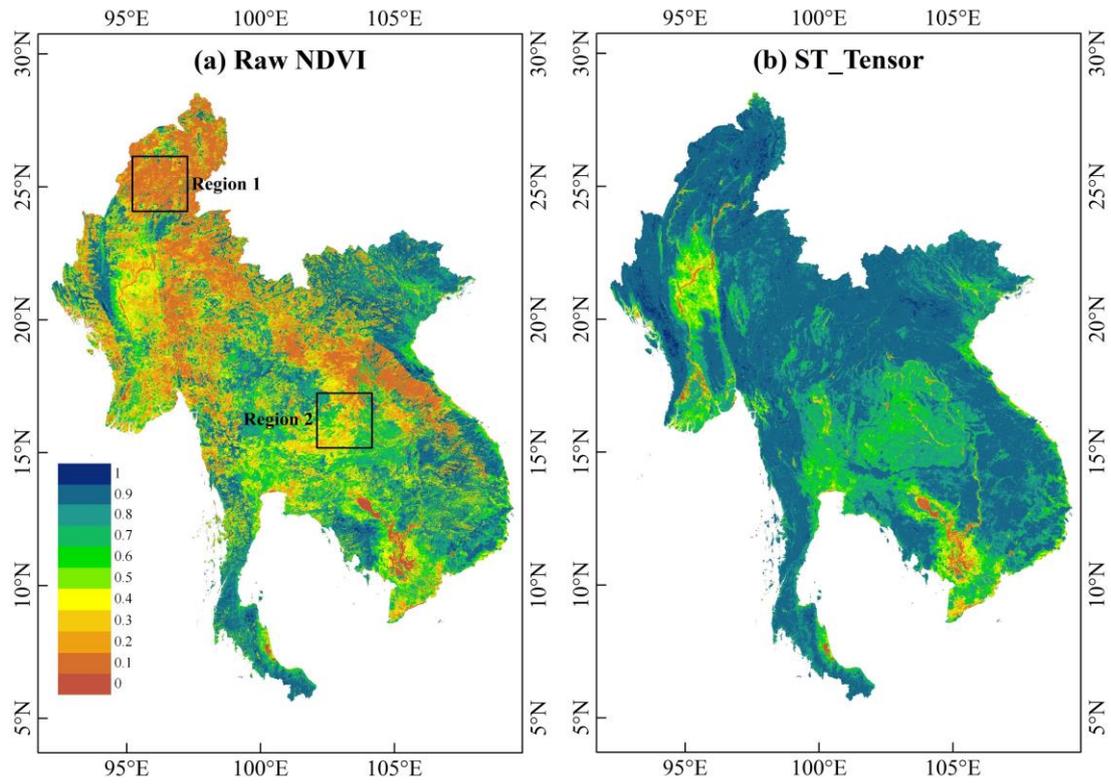

**Fig. 5.** The spatial performance of the proposed ST_Tensor method on DOY 241 in 2004 in the area of Mainland Southeast Asia.

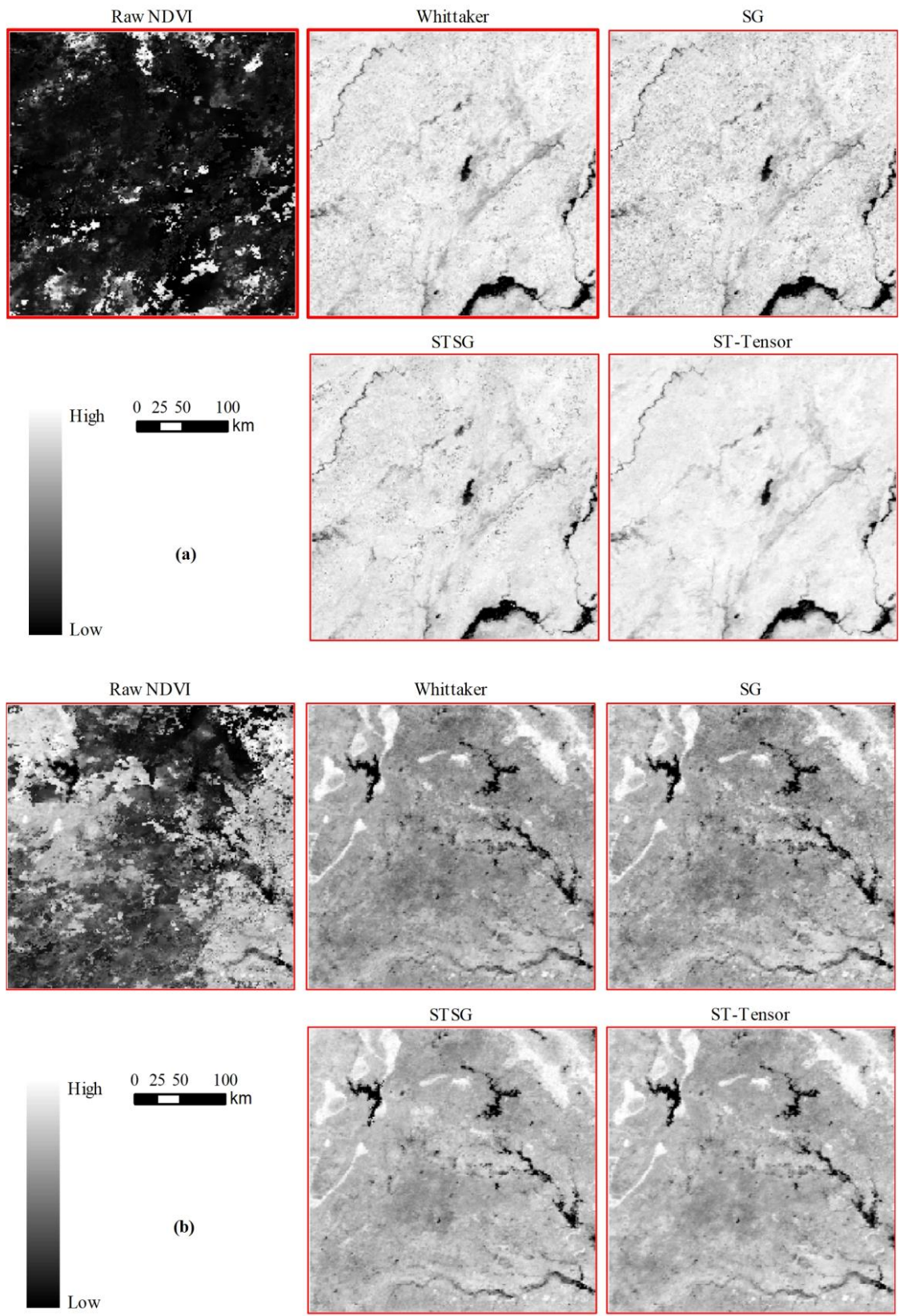

**Fig. 6.** The regional performance of the various methods in (a) Region 1, and (b) Region 2 in Fig.4.

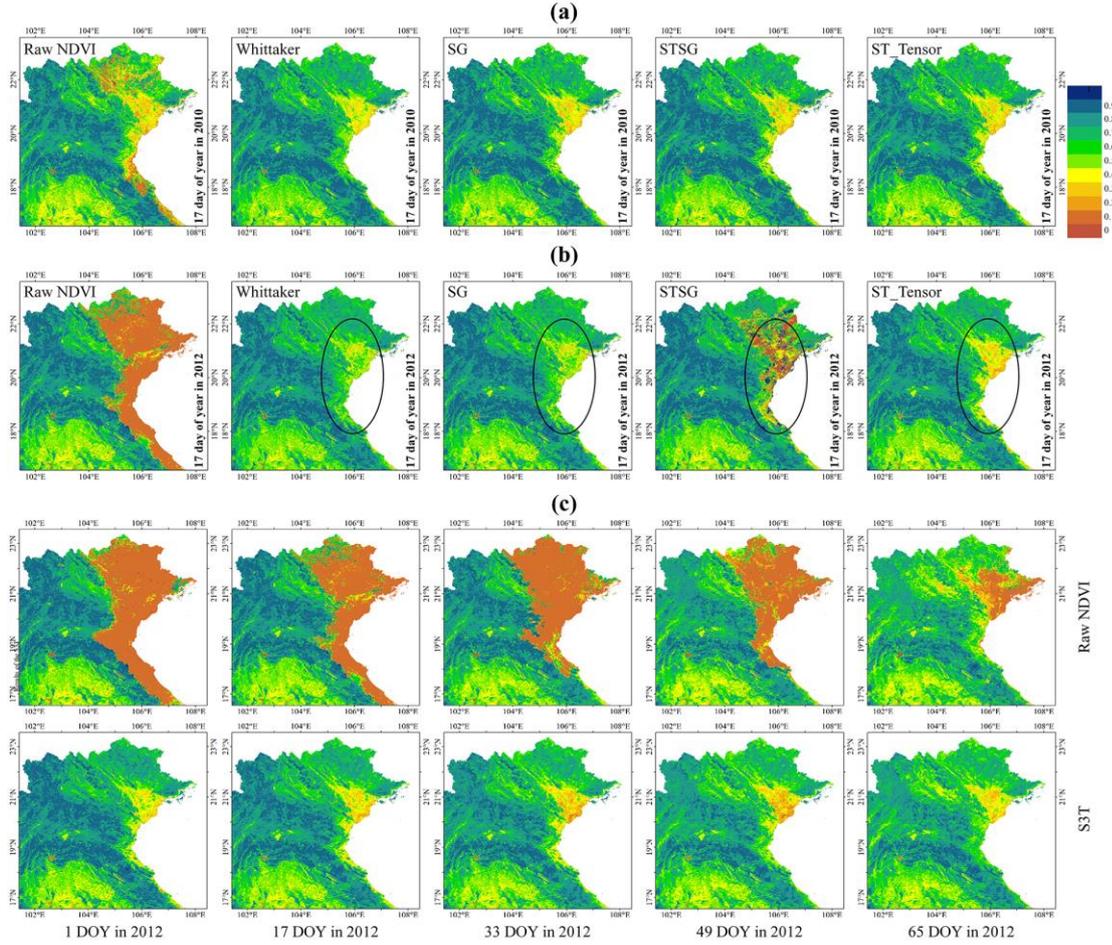

**Fig. 7.** The regional performance in northern Vietnam for (a) the various methods on DOY 17 in 2010, (b) DOY 17 in 2012, and (c) the proposed ST-Tensor method in the situation of spatio-temporally continuous missing data.

### 4.3 Quantitative assessment based on reference NDVI

To quantitatively evaluate the performance of the ST-Tensor method, a simulated experiment was conducted for the whole study area, and the quantitative indicator of the mean absolute error (MAE) was calculated. The simulated data were generated by calculating the reference NDVI, following the approach used in previous studies (Cao et al., 2018; Liu et al., 2017; Zhou et al., 2016). Since the true NDVI value of every pixel was unavailable, the reference NDVI was regarded as the true data, to obtain the quantitative metrics. The reference NDVI curve of a pixel is essentially a one-year time-series and can represent the vegetation growth pattern. It is calculated by averaging all the good NDVI values on the same day of the year for all years. Specifically, in this study, one pixel had 18 points for every day of the year during 2001–2018. If the number of good points was no less than 4 in all 18 points, then the average value of all the good points was calculated and regarded as the reference NDVI value. If the number of good points was less than 4 in all 18 points, then the corresponding reference NDVI value was calculated by linear interpolation. In this way, the resulting reference NDVI curve for each pixel was guaranteed to have little noise. The real RI values and the generated reference NDVI were then combined to create simulated contaminated data by adding gaps and noise. Specifically, if the RI value was equal to 0, the corresponding NDVI value was equal to the reference NDVI value; if the RI value was equal to 1, the corresponding NDVI value was equal to 95% of the reference NDVI value; if the RI value was equal to −1 or 3, the corresponding NDVI value was missing. By applying the different methods, the reconstruction results could be obtained and the MAE values could be calculated.

Fig. 8 shows the quantitative assessment results of the different methods in the whole study area, based on the simulated experiments, and Table 1 lists the corresponding statistics of the MAE values for the different methods. The Whittaker method performs the worst, with the largest mean MAE and the largest percentage of pixels showing an MAE of greater than 0.025, due to over-smoothing. The HANTS, IDR, and SG methods perform slightly better, in general. Comparing all the temporal methods, the SG method shows the best performance, but it is still inferior to the spatio-temporal methods, i.e., the STSG method and the ST-Tensor method. The proposed ST-Tensor method performs the best of all, and is also the most robust method. It obtains the lowest mean MAE of 0.012, the largest percentage of pixels showing an MAE of less than 0.01, and the lowest percentage of pixels having an MAE of more than 0.025. The success of the ST-Tensor method can be attributed to the comprehensive and integrated exploitation of the spatio-temporal information.

However, there are some local regions with large MAE values in the result of the ST-Tensor method. According to Fig. 1, it can be found that these regions have larger invalid data percentages. This shows that large numbers of missing data inevitably result in uncertain reconstructed values.

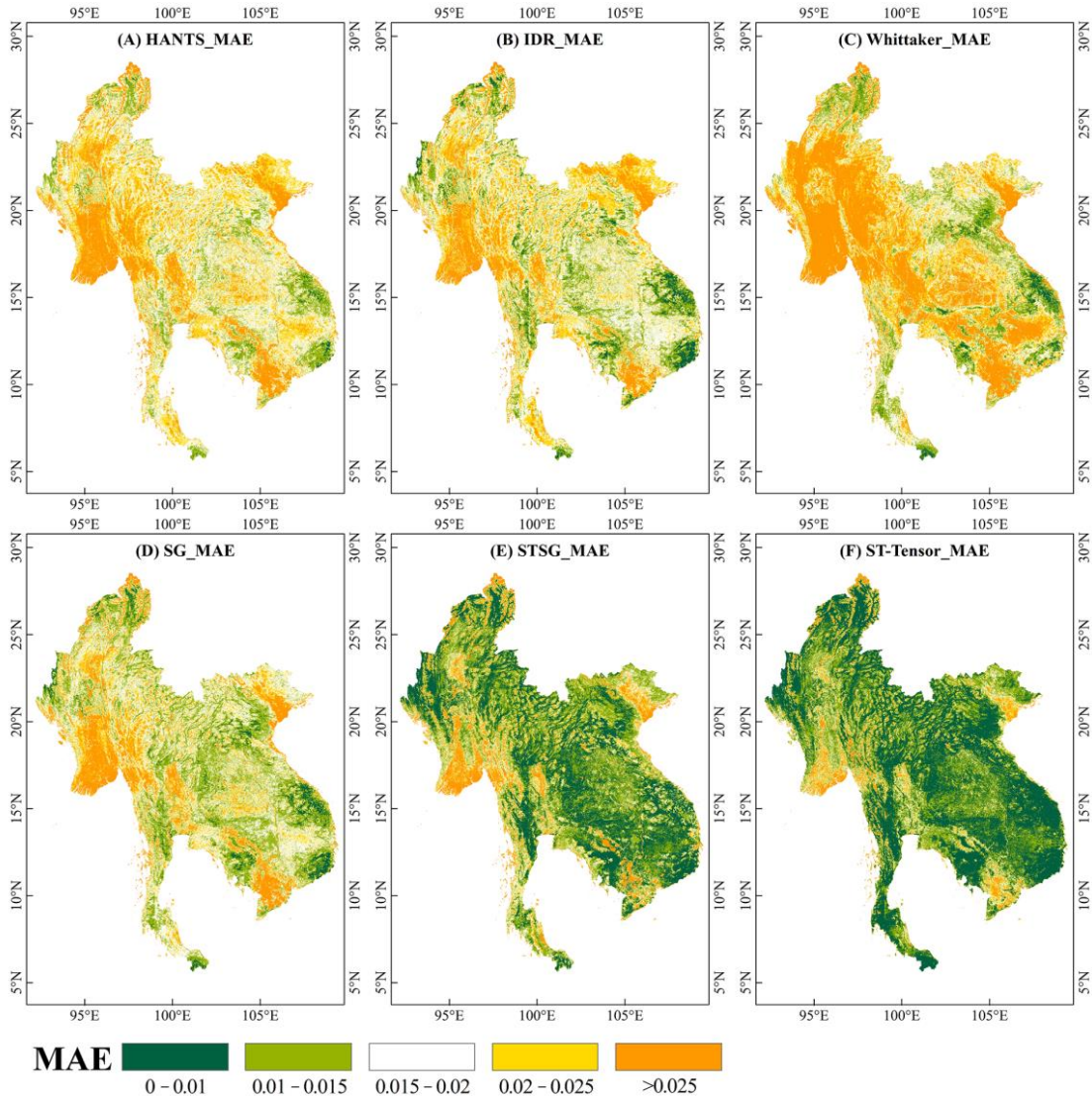

**Fig. 8.** The MAE of the various methods for simulated gaps and noise in the whole study area.

Table 1 The mean MAE values and the percentage for different MAE levels for the different methods, based on Fig. 11.

| MAE | HANTS | IDR | Whittaker | SG | STSG | ST-Tensor |
|---|---|---|---|---|---|---|
| Mean | 0.022 | 0.020 | 0.026 | 0.0195 | 0.016 | 0.012 |
| 0–0.01 | 0.6% | 3.6% | 2.1% | 3.8% | 33.0% | 47.2% |
| 0.01–0.015 | 14.3% | 21.5% | 16.6% | 31.0% | 35.4% | 33.0% |
| 0.015–0.02 | 38.6% | 36.6% | 23.8% | 34.2% | 14.4% | 10.3% |
| 0.02–0.025 | 25.0% | 21.2% | 19.6% | 14.7% | 6.8% | 4.4% |
| >0.025 | 21.5% | 17.1% | 37.9% | 16.2% | 10.3% | 5.0% |

## 5. Discussion

### 5.1 The gap-filling ability of the ST-Tensor method

Some recent experiments have proved that for Landsat time-series data, it is of great importance to the fill missing values before fitting the time series (Cao et al., 2020; Yan and Roy, 2020). This insight should also be true for low-resolution NDVI time-series data. However, most of the existing temporal reconstruction methods for NDVI time series use simple linear interpolation in the temporal neighborhood as a preprocessing step to fill the missing values before implementing temporal filtering (Chen et al., 2004; Liu et al., 2017). An important motivation for this study is that it is difficult for temporal linear interpolation to achieve precise reconstruction of contaminated pixels, especially in severely cloud-contaminated regions. Therefore, it was necessary to directly compare the reconstruction accuracy of the proposed ST-Tensor method with that of the linear interpolation method. We therefore undertook simulation experiments to illustrate the gap-filling performance of the ST-Tensor method.

Specifically, a sample patch with a spatial size of 400 × 400 (pixels) was selected in the study area. The missing data rate of the original sample patch in the whole 18-year time series was 22.2%. The following two scenarios were considered. Scenario 1: random gaps. Different proportions of random gaps were added to the original data. The total missing data rate ranged from 25% to 80% at 5% intervals; Scenario 2: spatio-temporally continuous gaps. A square area of size 50 × 50 (pixels) was fixed in the sample patch. This square area was then simulated as a continuous gap in the time series. The gap length, i.e., the number of temporally continuous missing observations, was varied from 2 to 12.

Fig. 9(a) shows the MAE results for the simulated random gaps at different missing data rates. By comparing the MAE, the better reconstruction ability of the ST-Tensor method is clear. In addition, with the increase of the missing data rate, the MAE of the ST-Tensor method increases linearly, with a gentle slope, while the MAE of the linear interpolation method increases exponentially. Fig. 9(b) presents the MAE results for the simulated spatio-temporally continuous gaps in different gap lengths, where the ST-Tensor method is again clearly better than the linear interpolation method. What is more, with the increase of the gap length, the MAE of the interpolation method increases more rapidly. When the gap length reaches the extreme value of 12, the MAE of the interpolation method is more than 0.09, which means that the reconstruction results have large errors. In contrast, the MAE of the ST-Tensor method is still around 0.03.

These results suggest that the ST-Tensor method not only obtains a higher reconstruction accuracy, but it is also more robust than the simple linear interpolation method. It can provide a high-quality gap-filling result, for both random and spatio-temporally continuous gaps.

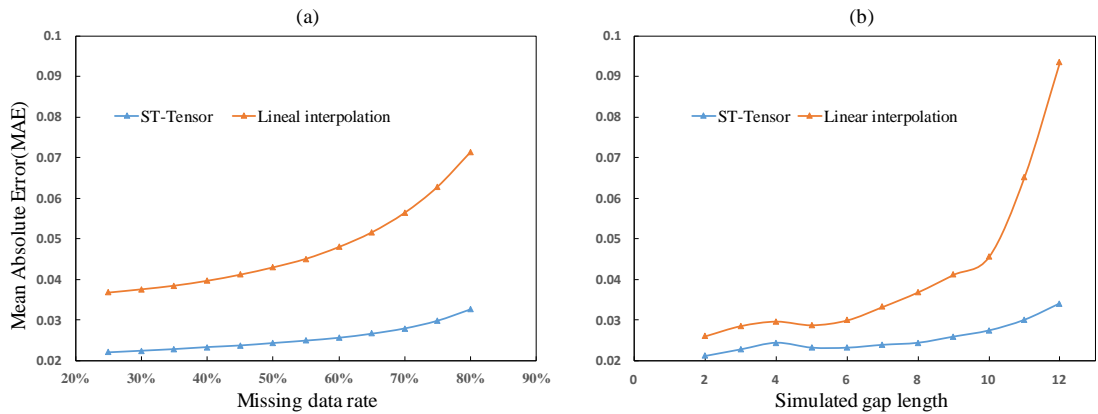

**Fig. 9.** Comparison of the mean absolute error (MAE) for the proposed ST-Tensor method (using the optimal parameters) and the simple linear interpolation method while (a) changing the missing data rate, and (b) changing the gap length.

## 5.2 The importance of tensor form rearrangement

In the procedure of tensor completion, we reshape the original tensor into a new tensor in order to make better use of the spatial neighborhood information and take the periodic temporal information into consideration. To prove the importance of this step, we undertook simulation experiments using the original tensor form and the proposed tensor form respectively, and compared the missing data reconstruction accuracy. Fig. 10(a) shows the results for simulated random gaps with missing data rates from 25% to 80% at a 5% interval. As can be seen, the MAEs for the proposed tensor form are slightly lower than for the original tensor form. However, in general, the two tensor forms show nearly the same performance in the case of random gaps, and the MAEs are in the range of 0.02 to 0.035. Fig. 10(a) provides the results for the simulated spatio-temporally continuous gaps with gap length from 2 to 12., which shows a huge difference in performance between the two tensor forms. That is, the MAEs for the original tensor form are more than 10 times those for the proposed tensor form, which are still in the range of 0.02 to 0.035.

Two things can be gleaned from these results. Firstly, low-rank tensor completion is an effective tool when filling gaps in time series NDVI images, and it can reach a higher reconstruction accuracy with the proper tensor form. Secondly, tensor rearrangement to account for the periodicity of the time series makes great sense. The original tensor form can express the spatial neighborhood information and temporal neighborhood information, which leads to its failure in filling spatio-temporally continuous gaps, because there is not enough valid information. The success of the proposed tensor form in reconstructing spatio-temporally continuous missing data can be attributed to the consideration of the periodic temporal information.

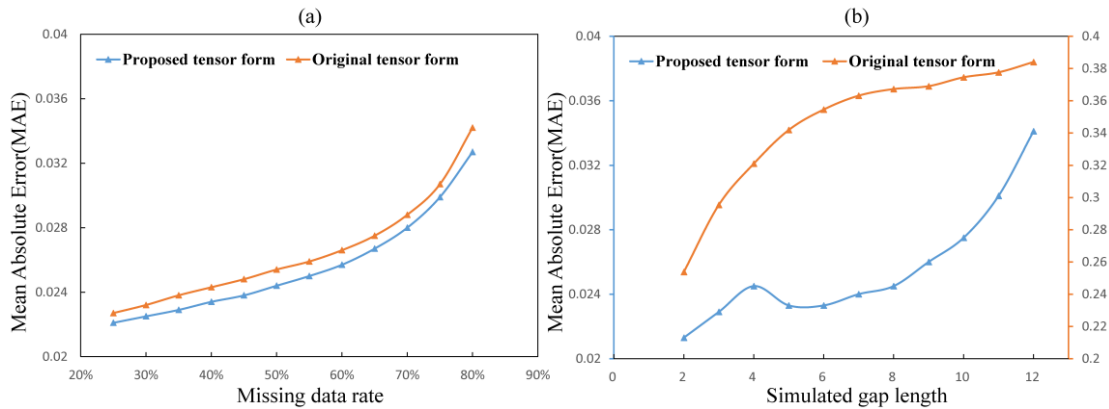

**Fig. 10.** Comparison of the mean absolute error (MAE) for (a) the proposed tensor form, and (b) the original tensor form.

**5.3 Parameter sensitivity of the ST-Tensor method**

The ST-Tensor method executes spatio-temporal tensor completion in a limited spatial patch, the size of which determines the utilization of the spatial information and further affects the low-rank reconstruction results. In general, the smaller the spatial patch, the stronger the spatial similarity, but the smaller the amount of spatial information that can be utilized. The bigger the spatial patch, the larger the amount of spatial information that can be used, but the weaker the spatial similarity. Therefore, the optimal size of a spatial patch can be obtained when the spatial similarity and the amount of spatial information reach a balance.

Fig. 11 shows the results for the simulated random gaps, from which a pattern can be clearly observed. When the missing data rate remains unchanged, as the spatial patch size increases, the MAE falls at first and then increases. Therefore, the optimal spatial size can be obtained according to the minimum MAE. It can be observed that when the missing data rate is larger, the corresponding optimal spatial patch size also tends to be larger, with a general range from 8 to 12. Although the optimal spatial patch size seems not to be fixed for the different missing data rates in this experiment, the fixed optimal patch size of 8 can be deemed as suitable for almost all conditions. The results suggest that the patch size should be set as 8 for a missing data rate of less than 60%, and only in the condition of a missing data rate of more than 65% should the patch size be more than 8, but such large patch size is rarely encountered in actual data. As a result, it can be concluded that the proposed method is not sensitive to the patch size for data with different missing data rates, and this parameter can be set as 8 in almost all conditions. As the missing data rate in the whole series was about 24%, we used a spatial patch size of 8 in all our experiments.

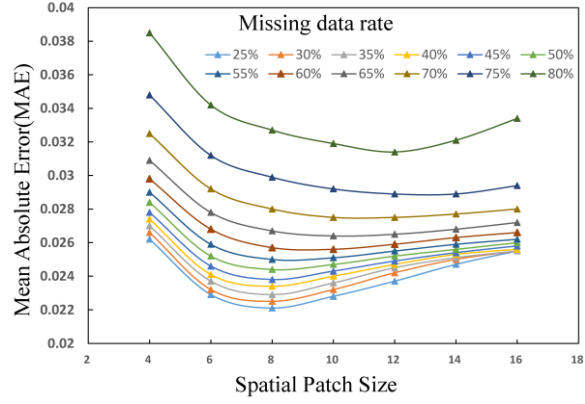

**Fig. 11.** The mean absolute error (MAE) for the proposed tensor completion method in different spatial patch sizes while changing the missing data rate.

**5.4 Uncertainties of the ST-Tensor method**

In this paper, an innovative ST-Tensor method has been proposed on the basis of low-rank tensor technology, which can effectively deal with the NDVI construction challenge in cloud-prone regions. We believe that this is the first time that low-rank tensor completion has been introduced into the field of NDVI reconstruction, as well as being the first time that $\ell_1$ trend filtering has been employed for NDVI noise reduction. The proposed method can synthetically use the multi-dimensional spatio-temporal information from the spatial neighbors, inter-annual variations, and periodic temporal characteristics, and thus has the ability to handle the challenge of missing key points (i.e., maximum or minimum NDVI points), temporally continuous gaps, or even spatio-temporally continuous gaps in NDVI data from rainy areas such as tropical and subtropical regions.

Although the ST-Tensor method shows a satisfactory ability to reconstruct time-series NDVI products, two uncertain issues remain that need to be considered. The first is that the proposed method uses a pixel reliability dataset as auxiliary data to distinguish valid from invalid pixels, as many other methods do. Thus, the uncertainty existing in the pixel reliability dataset may result in uncertainty of the reconstruction results. There are two scenarios: 1) Valid pixels are incorrectly marked as contaminated pixels. The reconstructed values of these pixels by tensor completion should be only slightly different from the true values, as illustrated in Section 5.1. Therefore, this uncertainty has little impact on the final results. 2) Contaminated pixels are incorrectly marked as valid pixels. In this scenario, the low values of these pixels would be retained after tensor completion. However, they would be processed by the following filtering procedure. As a result, the abrupt low values can be corrected. As a consequence, uncertainties in the pixel reliability dataset have only a limited impact on the final results.

The second concern is about pixels with land-cover changes, which is a common problem for almost all the methods. The performance of the proposed method on pixels with land-cover changes was tested in two typical cases, as shown in Fig. 12. It can be observed that the ST-Tensor method shows a good reconstruction ability in both cases. The results of the proposed method and the temporal methods, including the SG and Whittaker methods, can still simulate the processes of land-cover changes, but the STSG method shows some abnormal results. Although it is difficult to comprehensively conclude that the ST-Tensor method is not sensitive to land-cover changes, it has shown its stability and potential to deal with this issue.

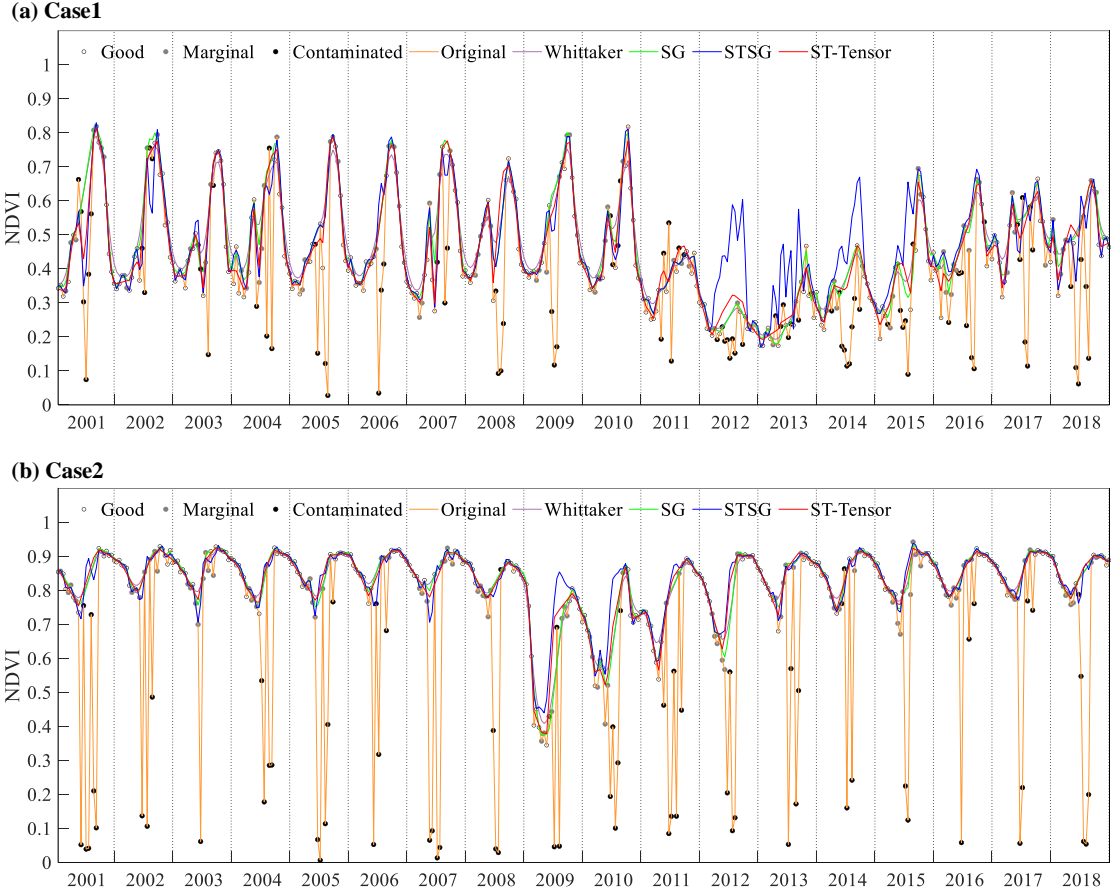

Fig. 12. Comparison of the results obtained by the Whittaker, SG, STSG, and ST-Tensor methods for two land-cover change cases.

## 6. Conclusions

The inevitable gaps and noise greatly hinder the wide application of NDVI time series, and this problem is particularly challenging in cloud-prone regions, due to the large percentages of missing data. In this paper, a new ST-Tensor method has been proposed to handle this problem, and we have described how the proposed method was applied to obtain a long-term NDVI product with a high quality for Mainland Southeast Asia. Spatio-temporal information from the spatial neighborhood, inter-annual variations, and periodic characteristics is fully considered in the shaped tensor form and effectively utilized with a weighted low-rank tensor completion technique. $\ell_1$ trend filtering is innovatively applied to further filter the NDVI time series, and its effectiveness and robustness have been proved in keeping the key feature points. The experiments showed that the proposed method outperformed five other widely used methods (the HANTS, IDR, Whittaker, SG, and STSG methods): 1) the ST-Tensor method showed a more powerful ability to deal with temporally continuous gaps and spatio-temporally continuous gaps; and 2) the ST-Tensor method also showed a good ability to preserve the key points in the seasonal variation curves, i.e., the peaks and valleys, and track the seasonal trajectory of NDVI. The time-series reconstruction results obtained for Mainland Southeast Asia showed that the proposed ST-Tensor method has the ability to deal with the NDVI gap-filling problem in cloud-prone areas. We expect that the proposed ST-Tensor method could be used to produce high-quality long-term NDVI products for almost all areas of the world, for the purpose of promoting vegetation-related ecological and environmental research.

## Acknowledgments

This research was supported by the National Key Research and Development Program of China (2019YFB2102903), and the National Natural Science Foundation of China (42001371).

## References

Atzberger, C., Eilers, P.H.C., 2011. A time series for monitoring vegetation activity and phenology at 10-daily time steps covering large parts of South America. Int. J. Digit. Earth 4, 365–386. https://doi.org/10.1080/17538947.2010.505664

Beck, P.S.A., Atzberger, C., Høgda, K.A., Johansen, B., Skidmore, A.K., 2006. Improved monitoring of vegetation dynamics at very high latitudes: A new method using MODIS NDVI. Remote Sens. Environ. 100, 321–334. https://doi.org/10.1016/j.rse.2005.10.021

Cao, R., Chen, J., Shen, M., Tang, Y., 2015. An improved logistic method for detecting spring vegetation phenology in grasslands from MODIS EVI time-series data. Agric. For. Meteorol. 200, 9–20. https://doi.org/https://doi.org/10.1016/j.agrformet.2014.09.009

Cao, R., Chen, Y., Chen, J., Zhu, X., Shen, M., 2020. Thick cloud removal in Landsat images based on autoregression of Landsat time-series data. Remote Sens. Environ. 249, 112001. https://doi.org/10.1016/j.rse.2020.112001

Cao, R., Chen, Y., Shen, M., Chen, J., Zhou, J., Wang, C., Yang, W., 2018. A simple method to improve the quality of NDVI time-series data by integrating spatiotemporal information with the Savitzky-Golay filter. Remote Sens. Environ. 217, 244–257. https://doi.org/10.1016/j.rse.2018.08.022

Chen, J., Jönsson, P., Tamura, M., Gu, Z., Matsushita, B., Eklundh, L., 2004. A simple method for reconstructing a high-quality NDVI time-series data set based on the Savitzky-Golay filter. Remote Sens. Environ. 91, 332–344. https://doi.org/10.1016/j.rse.2004.03.014

Chen, J.M., Deng, F., Chen, M., 2006. Locally adjusted cubic-spline capping for reconstructing seasonal trajectories of a satellite-derived surface parameter. IEEE Trans. Geosci. Remote Sens. 44, 2230–2237. https://doi.org/10.1109/TGRS.2006.872089

De Oliveira, J.C., Epiphanio, J.C.N., Rennó, C.D., 2014. Window regression: A spatial-temporal analysis to estimate pixels classified as low-quality in MODIS NDVI time series. Remote Sens. 6, 3123–3142. https://doi.org/10.3390/rs6043123

Didan, K., Munoz, A.B., Solano, R., Huete, A., 2015. MODIS Vegetation Index User's Guide (Collection 6) 2015, 31.

Dong, J., Xiao, X., Sheldon, S., Biradar, C., Duong, N.D., Hazarika, M., 2012. A comparison of forest cover maps in Mainland Southeast Asia from multiple sources: PALSAR, MERIS, MODIS and FRA. Remote Sens. Environ. 127, 60–73. https://doi.org/10.1016/j.rse.2012.08.022

Eilers, P.H.C., 2003. A perfect smoother. Anal. Chem. 75, 3631–3636. https://doi.org/10.1021/ac034173t

Gandy, S., Recht, B., Yamada, I., 2011. Tensor completion and low-n-rank tensor recovery via convex optimization.

Gerber, F., De Jong, R., Schaepman, M.E., Schaepman-Strub, G., Furrer, R., 2018. Predicting Missing Values in Spatio-Temporal Remote Sensing Data. IEEE Trans. Geosci. Remote Sens. 56, 2841–2853. https://doi.org/10.1109/TGRS.2017.2785240

Gu, J., Li, X., Huang, C., Okin, G.S., 2009. A simplified data assimilation method for reconstructing time-series MODIS NDVI data. Adv. Sp. Res. 44, 501–509. https://doi.org/10.1016/j.asr.2009.05.009

Guan, X., Shen, H., Li, X., Gan, W., Zhang, L., 2019. A long-term and comprehensive assessment of the urbanization-induced impacts on vegetation net primary productivity. Sci. Total Environ. 669, 342–352. https://doi.org/https://doi.org/10.1016/j.scitotenv.2019.02.361

Guan, X., Shen, H., Li, X., Gan, W., Zhang, L., 2018. Climate Control on Net Primary Productivity in the Complicated Mountainous Area: A Case Study of Yunnan, China. IEEE J. Sel. Top. Appl. Earth Obs. Remote Sens. 11, 4637–4648. https://doi.org/10.1109/JSTARS.2018.2863957

Huete, A., Didan, K., Miura, T., Rodriguez, E.P., Gao, X., Ferreira, L.G., 2002. Overview of the radiometric and biophysical performance of the MODIS vegetation indices. Remote Sens. Environ. 83, 195–213. https://doi.org/https://doi.org/10.1016/S0034-4257(02)00096-2

Ji, T., Huang, T., Zhao, X., Ma, T., Deng, L., 2017. A non-convex tensor rank approximation for tensor completion. Appl. Math. Model. 48, 410–422. https://doi.org/10.1016/j.apm.2017.04.002

Jönsson, P., Eklundh, L., 2002. Seasonality extraction by function fitting to time-series of satellite sensor data. IEEE Trans. Geosci. Remote Sens. 40, 1824–1832. https://doi.org/10.1109/TGRS.2002.802519

Julien, Y., Sobrino, J.A., 2010. Comparison of cloud-reconstruction methods for time series of composite NDVI data. Remote Sens. Environ. 114, 618–625. https://doi.org/10.1016/j.rse.2009.11.001

Kim, S.J., Koh, K., Boyd, S., Gorinevsky, D., 2009. ℓ 1 Trend filtering. SIAM Rev. 51, 339–360. https://doi.org/10.1137/070690274

Kolda, T.G., Bader, B.W., 2009. Tensor Decompositions and Applications ∗ 51, 455–500.

Kong, D., Zhang, Y., Gu, X., Wang, D., 2019. A robust method for reconstructing global MODIS EVI time series on the Google Earth Engine. ISPRS J. Photogramm. Remote Sens. 155, 13–24. https://doi.org/10.1016/j.isprsjprs.2019.06.014

Leinenkugel, P., Kuenzer, C., Dech, S., 2013a. Comparison and enhancement of MODIS cloud mask products for Southeast Asia. Int. J. Remote Sens. 34, 2730–2748. https://doi.org/10.1080/01431161.2012.750037

Leinenkugel, P., Kuenzer, C., Oppelt, N., Dech, S., 2013b. Characterisation of land surface phenology and land cover based on moderate resolution satellite data in cloud prone areas - A novel product for the Mekong Basin. Remote Sens. Environ. 136, 180–198. https://doi.org/10.1016/j.rse.2013.05.004

Liu, J., Musialski, P., Wonka, P., Ye, J., Member, S., 2013. Tensor Completion for Estimating Missing Values in Visual Data. IEEE Trans. Pattern Anal. Mach. Intell. 35, 208–220. https://doi.org/10.1109/TPAMI.2012.39

Liu, R., Shang, R., Liu, Y., Lu, X., 2017. Global evaluation of gap-filling approaches for seasonal NDVI with considering vegetation growth trajectory, protection of key point, noise resistance and curve stability. Remote Sens. Environ. 189, 164–179. https://doi.org/10.1016/j.rse.2016.11.023

Lu, X., Liu, R., Liu, J., Liang, S., 2007. Removal of noise by wavelet method to generate high quality temporal data of terrestrial MODIS products. Photogramm. Eng. Remote Sensing 73, 1129–1139. https://doi.org/10.14358/PERS.73.10.1129

Lunetta, R.S., Knight, J.F., Ediriwickrema, J., Lyon, J.G., Worthy, L.D., 2006. Land-cover change detection using multi-temporal MODIS NDVI data. Remote Sens. Environ. 105, 142–154. https://doi.org/https://doi.org/10.1016/j.rse.2006.06.018

Moreno, Á., García-Haro, F.J., Martínez, B., Gilabert, M.A., 2014. Noise reduction and gap filling of fAPAR time series using an adapted local regression filter. Remote Sens. 6, 8238–8260. https://doi.org/10.3390/rs6098238

Myneni, R.B., Williams, D.L., 1994. On the relationship between FAPAR and NDVI. Remote Sens. Environ. 49, 200–211. https://doi.org/https://doi.org/10.1016/0034-4257(94)90016-7

Piao, S., Liu, Q., Chen, A., Janssens, I.A., Fu, Y., Dai, J., Liu, L., Lian, X., Shen, M., Zhu, X., 2019. Plant phenology and global climate change: Current progresses and challenges. Glob. Chang. Biol. 25, 1922–1940. https://doi.org/10.1111/gcb.14619

Shao, Y., Lunetta, R.S., Wheeler, B., Iiames, J.S., Campbell, J.B., 2016. An evaluation of time-series smoothing algorithms for land-cover classifications using MODIS-NDVI multi-temporal data. Remote Sens. Environ. 174, 258–265. https://doi.org/https://doi.org/10.1016/j.rse.2015.12.023

Shen, H., Li, X., Cheng, Q., Zeng, C., Yang, G., Li, H., Zhang, L., 2015. Missing Information Reconstruction of Remote Sensing Data: A Technical Review. IEEE Geosci. Remote Sens. Mag. 3, 61–85. https://doi.org/10.1109/MGRS.2015.2441912

Sloan, S., Jenkins, C.N., Joppa, L.N., Gaveau, D.L.A., Laurance, W.F., 2014. Remaining natural vegetation in the global biodiversity hotspots. Biol. Conserv. 177, 12–24. https://doi.org/10.1016/j.biocon.2014.05.027

Tang, X., Bullock, E.L., Olofsson, P., Estel, S., Woodcock, C.E., 2019. Near real-time monitoring of tropical forest disturbance: New algorithms and assessment framework. Remote Sens. Environ. 224, 202–218. https://doi.org/https://doi.org/10.1016/j.rse.2019.02.003

Tucker, C.J., Sellers, P.J., 1986. Satellite remote sensing of primary production. Int. J. Remote Sens. 7, 1395–1416. https://doi.org/10.1080/01431168608948944

Verger, A., Baret, F., Weiss, M., Kandasamy, S., Vermote, E., 2013. The CACAO method for smoothing, gap filling, and characterizing seasonal anomalies in satellite time series. IEEE Trans. Geosci. Remote Sens. 51, 1963–1972. https://doi.org/10.1109/TGRS.2012.2228653

VIOVY, N., ARINO, O., BELWARD, A.S., 1992. The Best Index Slope Extraction ( BISE): A method for reducing noise in NDVI time-series. Int. J. Remote Sens. 13, 1585–1590. https://doi.org/10.1080/01431169208904212

Xiao, X., Boles, S., Frolking, S., Li, C., Babu, J.Y., Salas, W., Moore, B., 2006. Mapping paddy rice agriculture in South and Southeast Asia using multi-temporal MODIS images. Remote Sens. Environ. 100, 95–113. https://doi.org/10.1016/j.rse.2005.10.004

Xu, L., Li, B., Yuan, Y., Gao, X., Zhang, T., 2015. A temporal-spatial iteration method to reconstruct NDVI time series datasets. Remote Sens. 7, 8906–8924. https://doi.org/10.3390/rs70708906

Yan, L., Roy, D.P., 2020. Spatially and temporally complete Landsat reflectance time series modelling: The fill-and-fit approach. Remote Sens. Environ. 241. https://doi.org/10.1016/j.rse.2020.111718

Yang, G., Shen, H., Zhang, L., He, Z., Li, X., 2015. A moving weighted harmonic analysis method for reconstructing high-quality SPOT VEGETATION NDVI time-series data. IEEE Trans. Geosci. Remote Sens. 53, 6008–6021. https://doi.org/10.1109/TGRS.2015.2431315

Zhang, X., Friedl, M.A., Schaaf, C.B., Strahler, A.H., Hodges, J.C.F., Gao, F., Reed, B.C., Huete, A., 2003. Monitoring vegetation phenology using MODIS. Remote Sens. Environ. 84, 471–475. https://doi.org/https://doi.org/10.1016/S0034-4257(02)00135-9

Zhou, J., Jia, L., Menenti, M., 2015. Reconstruction of global MODIS NDVI time series: Performance of Harmonic ANalysis of Time Series (HANTS). Remote Sens. Environ. 163, 217–228. https://doi.org/10.1016/j.rse.2015.03.018

Zhou, J., Jia, L., Menenti, M., Gorte, B., 2016. On the performance of remote sensing time series reconstruction methods – A spatial comparison. Remote Sens. Environ. 187, 367–384. https://doi.org/10.1016/j.rse.2016.10.025

Zhu, W., Pan, Y., He, H., Wang, L., Mou, M., Liu, J., 2012. A changing-weight filter method for reconstructing a high-quality NDVI time series to preserve the integrity of vegetation phenology. IEEE Trans. Geosci. Remote Sens. 50, 1085–1094. https://doi.org/10.1109/TGRS.2011.2166965